\documentclass[manuscript]{aastex}

\shorttitle{Plasma Heating to Super-Hot Temperatures}
\shortauthors{Sharykin et al}

\begin{document}

\title{Plasma Heating to Super-Hot Temperatures ($>$30 MK) in the August 9, 2011 Solar Flare}

\author{Sharykin I.N., Struminsky A.B. and Zimovets I.V.}
\affil{Space Research Institute, Russian Academy of Science, Moscow 117997, Russia // 
}


\begin{abstract}
We investigate the August 9, 2011 solar flare of X-ray class X6.9, the ``hottest'' flare from 2000 to 2012, with a peak plasma temperature according to GOES data of $\approx$32.5 MK. Our goal is to determine the cause of such an anomalously high plasma temperature and to investigate the energy balance in the flare region with allowance made for the presence of a super-hot plasma ($>$30 MK). We analyze the RHESSI, GOES, AIA/SDO, and EVE/SDO data and discuss the spatial structure of the flare region and the results of our spectral analysis of its X-ray emission. Our analysis of the RHESSI X-ray spectra is performed in the one-temperature and two-temperature approximations by taking into account the emission of hot ($\sim$20 MK) and super-hot ($\sim$45 MK) plasmas. The hard X-ray spectrum in both models is fitted by power laws. The observed peculiarities of the flare are shown to be better explained in terms of the two-temperature model, in which the super-hot plasma is located at the flare loop tops (or in the magnetic cusp region). The formation of the super-hot plasma can be associated with its heating through primary energy release and with the suppression of thermal conduction. The anomalously high temperature (33 MK according to GOES) is most likely to be an artefact of the method for calculating the temperature based on two-channel GOES measurements in the one-temperature approximation applied to the emission of a multi-temperature flare plasma with a minor contribution from the low-temperature part of the differential emission measure.
\end{abstract}

\keywords{flares, X-ray emissions, accelerated electrons}

\section*{INTRODUCTION AND FORMULATION OF THE PROBLEM}

It is customary to judge the power of solar flares from the intensity of their soft X-ray (SXR) emission recorded in the 1—8 A channel by the GOES X-ray detector. The peak mean intensity recorded in 1 min is considered to be the X-ray class of a flare. The plasma temperature T and emission measure EM can be estimated in the one-temperature approximation from GOES measurements in two energy channels (the second channel is 0.5—4.0~\AA) (Thomas et al.~1985); the smaller the difference between the fluxes in both channels, the higher the temperature.

Ryan et al.~(2012) performed a statistical analysis of the SXR fluxes, temperatures, and emission measures for solar flares based on GOES data. It can be seen from their results that the relative scatter of flare temperatures is fairly small compared to the scatter of emission measures and T $>$30 MK are fairly rare; the events with temperatures above this arbitrary threshold can be classified as ``super-hot'' ones (Caspi et al.~2014). However, the GOES temperature estimate proceeds from the one-temperature approximation, which is by no means always valid. Spectral observations of the X-ray and ultraviolet (UV) emissions should be used to obtain more accurate information about the temperature distribution in the flare region.

Here, we investigate the unique GOES X6.9 solar flare occurred on August 9,~2011. The uniqueness of this event consists in a high plasma temperature, which is estimated from GOES data to be $\approx$32.5 MK (see the table). Such a temperature allows this event to be assigned to super-hot flares and, moreover, it is the hottest event among the M and X flares (Fig.\,1) from 2000 to 2012. However, the existence of high flare plasma temperatures can be firmly established only by means of a detailed spectral analysis (for example, the high temperature can be associated with the nonthermal emission that contributes to the short-wavelength GOES channel or the ``temperature'' is measured on the basis of RHESSI X-ray observations).

One of the first studies of the super-hot flare plasma was carried out on the basis of SMM observations by Svestka and Poletto~(1985), who discussed the relationship of the observed super-hot plasma to magnetic reconnection. Based on SMM data, Den and Somov~(1989) described the observations of super-hot points with temperatures of 50-60 MK. Kovalev et al.~(2001) investigated the super-hot structures with temperatures of 20—50 MK. observed from the Yohkoh spacecraft. The temperature determination from SMM and Yohkoh data in these papers was based on observations not in two channels, as with GOES, but in three or four. However, the nonthermal emission may have made a great contribution to the high-energy channels and the analysis was improper. The present-day RHESSI spacecraft has much better spatial, temporal, and spectral resolutions, which allows the flare plasma temperature to be analyzed more reliably.

A spectral analysis of the X-ray emission based on RHESSI data shows the presence of a plasma with a temperature of $\approx$40 MK in some flares. For example, Caspi and Lin~(2010) analyzed the X-ray spectrum of the July 23,~2002~(X4.8) flare in the two-temperature approximation and found the temperature of the hotter component to peak at $\approx$45 MK. A comparison of the July 23,~2002 and August 9,~2011 flares based on GOES data is made in the table. According to GOES data, the August 9,~2011 event is hotter than the July 23,~2002 event by 10 MK (Fig. 1). At such a difference in temperatures, it can be assumed that the contribution from the super-hot plasma to the X-ray flux in the August 9,~2011 event is more significant than that in the July~23,~2002 event. Liu et al.~2013) considered the short burst of X-ray emission (12—25 keV, RHESSI) in the B-class flare occurred on June 26, 2002, as a thermal one with a characteristic peak temperature of $\approx$36 MK. The statistical study by Caspi et al. (2014), where the possibility of dense coronal plasma heating to super-hot temperatures is discussed, was a continuation of the work by Caspi and Lin~(2010).

The presence of a super-hot plasma in the region of primary solar flare energy release was previously discussed by Somov and Kosugi~(1997) in their theoretical work. The fact that there is a super-hot plasma is important for investigating the primary flare energy release and particle acceleration, while the plasma of ``typical'' flare temperatures according to GOES data is probably associated with chromospheric ``evaporation'' (Fisher et al.~1985; Boiko and Livshits~1995), i.e., it is a secondary phenomenon with respect to the primary energy release.

\begin{table}[h!]
\begin{center}
	\begin{tabular}{|c|c|c|c|c|c|}
\hline
Flare (GOES class), & $I_{max}($1-8 \AA), & $I_{max}$(0.5-4 \AA), & $I_{4}/I_{8}$ & $T_{max}$, & $EM_{max}$, \\
Date and UT time  &  $10^{-4}$ W m$^{-2}$ &  $10^{-4}$ W m$^{-2}$ &  & MK & $10^{49}$ cm$^{-3}$ \\
\hline
August 9, 2011 (X6.9), 07:40 & 7.4 & 3.5 & 0.47 & 32.5 & 29.2 \\
July 23, 2002 (X4.8), 00:30 & 5.1 & 1.7 & 0.33 & 22.6 & 23.9 \\
\hline
\end{tabular}
\caption{Comparison of the August 9, 2011 (X6.9) and July 23, 2002 (X4.8) flares based on GOES observations}
\end{center}
\end{table}

The goal of our paper is to determine the cause of the anomalously high temperature determined from GOES data in the August 9, 2011 event and to investigate the energy balance in the flare region by taking into account the presence of a super-hot plasma with a temperature of more than 30 MK- Below, we will present: the GOES and RHESSI X-ray observations as well as the spatially resolved AIA/SDO extreme ultraviolet (EUV) observations (Section~1); our analysis of the RHESSI spectra in the two-temperature and one-temperature approximations (Section~2); our discussion of the possible cause of the anomalously high temperature according to GOES (Section~3); our discussion of the energy balance and electron acceleration in the flare region (Section~4); our discussion of the results obtained and conclusions (Section~5).

\section{OBSERVATIONS}

\subsection{Data}

We use observational data from the following instruments:

(1)	The GOES-15 (Geostationary Operational Environmental Satellite) soft X-ray detectors: the SXR observations in the short-wavelength (0.5—4~\AA) and long-wavelength (1—8~\AA) channels. The time resolution is 2 s. The GOES data are used to investigate the temporal plasma heating dynamics.

(2)	The Reuven Ramaty High Energy Solar Spectroscopic Imager (RHESSI) (Lin et al.~2002): the spatially resolved and spectral X-ray observations in a wide energy range, 3 keV—7 MeV. In our paper, we investigate the X-ray emission only in the range 3—300 keV to understand the spatial structure, plasma heating dynamics, and electron acceleration.

(3)	The Atmospheric Imaging Assembly (AIA) (Lemen et al.~2012) onboard the SDO (Solar Dynamics Observatory) spacecraft: the spatially resolved EUV observations (94, 131, 171, 193, 211, 304, and
335~\AA) and UV observations (1600 and 1700~\AA). The time resolution is 12 s and the spatial resolution is 1.2$^{\prime\prime}$ (0.6$^{\prime\prime}$ pixel size). In our paper, we use only the 94~\AA observations, because this channel has the lowest sensitivity and the CCD array undergoes lesser oversaturation during solar flares. The AIA data are used to investigate the spatial structure of the flare region.

(4)	The Extreme Ultraviolet Variability Experiment (EVE) (Woods et al.~2012) onboard the SDO spacecraft: the spectral UV observations of the entire solar disk. The time resolution is 10 s and the
wavelength range is 6.5-40.0 nm with a spectral resolution of 0.02 nm. The UV spectra are used to estimate the radiative losses of the flare region in the ultraviolet.

The Solar Software package (http://www.lmsal.com/solarsoft/) is used to analyze the observational data.

\subsection{Flare Observations}

Figure\,2 presents the time profiles for the RHESSI X-ray photon count rate and the X-ray flux from GOES data. The pre-impulsive phase starting approximately from 07:45:00 UT (Universal Time) is identified in the flare, but we do not consider it in this paper. The impulsive phase (the period in which HXR emission with an energy $>$50 keV is observed) begins approximately at 08:01:00 UT, lasts approximately until 08:06:00 UT, and is characterized by a sequence of several HXR bursts with the peak intensity at 08:02:08 UT. We investigate the processes of energy release in the flare region that occurred in the time interval 08:02:00-08:06:00 UT indicated in Fig.\,2 by the gray band. This time interval was chosen, because it is characterized by one position of the RHESSI attenuator and covers the bulk of the impulsive phase with HXR emission.

The flare region in the chosen event was close to the limb of the solar disk (N18W82). This allows us to investigate the vertical spatial structure, because the projection effect is minor (Fig.\,3). The AIA images in the 94~\AA channel show that an arcade of loops is involved in the flare process, as evidenced by the presence of two brightening bands, flare ribbons. The loop structure in the upper left corner of the image corresponds to the region of pre-impulsive flare energy release and is not the subject of our discussion in this paper. Despite the low sensitivity of the 94~\AA channel and the short exposure time, CCD pixel saturation, which is especially strong at 08:02:38 UT, is nevertheless observed in the lower part of the image. However, no X-ray sources are observed with RHESSI in this region and we do not consider it.

The chosen time intervals for imaging correspond to individual HXR bursts. The X-ray source evolve in the flare region inside the central part of the loop arcade. Two HXR sources in the energy range 60-200 keV were observed at the initial time 08:02:38 UT. They coincided in space with the flare ribbons observed by AIA (Fig.~ÇÀ) and, consequently, were located at the flare loop footpoints. At the same time, the SXR source was located at the flare loop top. Only one HXR source located above the flare ribbons and coinciding in space with the SXR source was observed at succeeding times (Figs.\,3B~and~3C). Both HXR and SXR sources were probably in the coronal part of the flare loops at these times.

The characteristic sizes of the SXR sources are estimated from the X-ray images to be $\approx$10$^{\prime\prime}$ or, given their latitude $\phi\approx$60 deg., approximately $8\times10^8$ cm. The ribbon separation is estimated from the UV images to be $\approx$6$^{\prime\prime}$, the size of individual bright points in the ribbons is $\approx$2$^{\prime\prime}$, the volume of the flare region is estimated to be $\approx 10^{27}$ cm$^{3}$, and this value will be used below for our calculations (Section~3).

\section{SPECTRAL ANALYSIS OF THE X-RAY EMISSION BASED ON RHESSI DATA}

For the event under consideration, we construct its spectra in four-second time intervals and perform its spectral analysis in the OSPEX package by the least-squares method. The X-ray spectra are investigated in the two-temperature and one-temperature approximations. This is because the spectrum in the range $\approx$25-50 keV has the characteristic shape of a wide cap (see Fig.\,4) and can be explained both by the thermal emission of a super-hot plasma and by the bremsstrahlung of nonthermal electrons with a soft spectrum. The X-ray spectrum above 50 keV is fitted by a power law and is assumed to be formed through the nonthermal bremsstrahlung of accelerated electrons. The results of our spectral analysis will be used in Section 3 to calculate the thermal energy of the plasma and the nonthermal energy of the accelerated electrons.

An example of the two-temperature fit to the X-ray spectrum is presented in Fig.\,4 (left panel). The thermal part ($<$50 keV) is fitted by two continuum spectra of one-temperature plasma emission with temperatures $T_h=$20 MK and $r_{sh}=$39 MK with allowance made for the set of Fe and Ni (6.3 and 8 keV) lines based on the CHIANTI model (Dere et al.~2009). The nonthermal part of the X-ray spectrum ($>$50 keV) is fitted by a double power law with a break at $Å_{br}=$20 keV and a power-law slope below this energy equal to 1.5. This break models the low-energy boundary of the nonthermal electron spectrum (Holman,~2003). The value of 20 keV was chosen arbitrarily so that the break was masked as the thermal part of the X-ray spectrum. As a result, we have six free parameters being varied in the least-squares method: the temperature $T_h$, the emission measure $EM_h$, $T_{sh}$, $EM_{sh}$, the normalization coefficient of the photon spectrum for an energy of 50 keV $A_{ph}$, and the power-law index of the X-ray spectrum for $E>E_{br}$ =20 keV $\gamma$. The indices h (hot) and sh (super-hot) denote the hot and super-hot plasmas, respectively.

The right part of Fig.\,4 shows an example of the one-temperature interpretation of the RHESSI X-ray spectrum. In this case, we use a one-temperature model of continuum emission with lines and a triple power-law fit of the HXR  spectrum. The first break $E_{low}$ models the low-energy boundary just as was done in the two-temperature approximation. The second break $E_{br}$ corresponds to joining near $\approx50$ keV between the soft and hard parts of the HXR spectrum. As a result, seven parameters are varied in the least-squares method: $EM$, $T$, $A_{ph}$, $E_{low}$, $E_{br}$, $\gamma_s$, and $\gamma_h$, where s and h stand for soft and hard, respectively.

The described model fits to the X-ray spectra are used in the entire time intervals indicated in Fig.\,2 by the gray band. Below, we use the results of our spectral analysis to calculate the energy release in the flare region.

\section{DISCUSSION OF THE CAUSE OF THE ANOMALOUSLY HIGH TEMPERATURE MEASURED WITH GOES}

Our spectral analysis of the RHESSI data shows the presence of a super-hot plasma ($T>$40 MK), with this value being comparable to the temperature of the super-hot component obtained for the July 23,~2002 flare (Caspi and Lin~2010). The emission measure of the super-hot plasma in this paper was found to be several times larger than that for the August 9,~2011 flare. However, the sensitivity of the GOES detectors to photons with energies above 20 keV is low (Fig.~5, the right panel) and, therefore, the super-hot plasma emission affects weakly the recorded X-ray fluxes and the temperature determined from GOES data.

The emission with energies below 20 keV, which is emitted to a greater extent by the hot plasma and not by the super-hot one, makes the largest contribution to the recorded (GOES) flux. Hence the following explanation of the anomalously high temperature measured with GOES can be suggested. Above, it has been pointed out that the technique for determining the temperature from GOES measurements is based on a comparison of the recorded X-ray fluxes in two channels under the assumption of a one-temperature plasma (Thomas et al.~1985). The smaller the difference between the X-ray intensities in the two GOES channels, the higher the temperature. Accordingly, the ``temperature'' can be increased through a powerful flux in the short-wavelength channel or through a weak flux in the long-wavelength one. It is the second variant that seems to be the most likely cause of the anomalously high temperature for the August 9,~2011 event, because the sensitivity of the GOES X-ray detectors to the emission of a plasma with a temperature above 40 MK is low.

A reduced flux in the short-wavelength GOES channel can be obtained by considering a multi-temperature plasma with the suppressed low-temperature component of the differential emission measure (DEM). This effect can be modeled in the following simplified way. Consider DEM in the form of an exponential (a rapid decrease in the amount of plasma with temperature):
$$
DEM(T) = EM_2\cdot exp\left(\frac{2-T}{T_{scale}}\right)\mbox{~  [cm}^{-3}\mbox{K}^{-1}\mbox{]}
$$
where $T_{scale}$ and $T$ have the dimensions [keV] and the normalization factor $EM_2$ is the emission measure of a plasma with a temperature of 2 keV. The chosen DEM has a low-temperature boundary $T_{low}$ and a high-temperature boundary $T_{high}$. The super-hot component contributes weakly to the flux recorded by the GOES X-ray detectors; therefore, $T_{high}$ is fixed at 4 keV and $T_{low}$ is a variable parameter. The left panel in Fig.\,5 presents the X-ray spectra modeled with the CHIANTI package (Dere et al.~2009); the abundances of chemical elements are assumed to be equal to the coronal ones. It can be seen that the low-energy part ($<$20 keV) of the X-ray spectrum in the case of small $T_{low}$ lies higher than in the case of large $T_{low}$. The spectra above 10 keV in both cases differ little from one another. Accordingly, the flux in the long-wavelength GOES channel will be lower for DEM with larger $T_{low}$; hence we will obtain a higher temperature according to the method of Thomas et al.~(1985).

Thus, we conclude that events with an anomalously high temperature according to GOES are characterized by a small amount of plasma at relatively low temperatures ($T<$20 MK) compared to ``normal'' flares. The presence of a super-hot plasma ($T>$30 MK) affects weakly the X-ray flux recorded by the GOES detectors. The possible nature of the small amount of low-temperature plasma is discussed in Section 5.

\section{ENERGY RELEASE IN THE FLARE REGION}

\subsection{A Model of the Flare Region in the Two-Temperature Approximation}

Let us discuss an approximate model of the flare region shown schematically in Fig.\,6 that describes the two-temperature fit of the RHESSI X-ray spectra. We assume that a region with a super-hot plasma with a temperature $T_{sh}$ and an emission measure $EM_{sh}$ (we will call it the sh region) is located in the upper part of the flare loop or in the magnetic cusp region (which is inessential for our model) and a region with a hot plasma ($T_h$, and $EM_h$, the h region) is located below. In the lower part of the loop, the up arrow indicates the plasma outflow into the coronal region (chromospheric evaporation).

The observed plasma in the sh region is assumed to be a source of nonthermal electrons and is connected with the region of their acceleration. This assumption is based on the fact that the sh region is nearest to the presumed place of primary energy release in the corona (magnetic reconnection). The produced population of nonthermal electrons is injected into the h region and subsequently into the dense solar atmosphere, leading to chromospheric evaporation. The wavy lines at the boundary of the sh region in Fig.\,6 indicate the connection with the region of primary energy release, which can have, for example, a cusp-like shape or a more complex geometry. We do not discuss the details of the process of primary energy release, because the observational data are limited.

In addition to the heating through the interaction of plasma particles with nonthermal electrons, we assume the presence of a heat flow from the sh region into the h region of the flare loops. In the model under consideration, all transport processes are assumed to occur along magnetic field lines. At the top of the loop and at its footpoints, the temperature gradient is taken to be zero (MacNeice et al.~1984; Boiko and Livshits~1995). Thus, a redistribution of heat occurs only inside the flare loop.

The distributions of centroid separations for the X-ray sources in two different energy channels (3-15 and 25-50 keV, 3-10 and 15-30 keV) in different time intervals are presented on the histogram (Fig.\,6). One of the ranges corresponds to the emission of the plasma with $T_h$, while the emission of the plasma with $T_{sh}$ is concentrated in the other energy range. On average, the centroid separation, given the dispersion, is $\approx0.5±0.3$ of the $2^{\prime\prime}\times 2^{\prime\prime}$ RHESSI pixel. This result provides evidence for the fact that the sh region lies slightly higher than the h region and confirms the chosen geometry of the flare region.

The AIA UV images show that not one loop but a whole arcade of loops is involved in the flare process. However, the described model of the processes in the flare region for one loop is extended to the part of the loop arcade where the SXR and HXR emissions are observed. We assume that, on average, the thermodynamic and geometric parameters of the loops in this part of the arcade are the same. If there were a great difference in temperature from one loop to another, then we would observe a displacement of the X-ray sources with different energies along the arcade, which, however, is not observed. The loops with a low emission measure, accordingly, make a smaller contribution to the overall energetics.

\subsection{Calculating the Plasma Internal Energy}

The internal energy of a fully ionized hydrogen plasma is calculated from the emission measure, temperature, and volume of the flare region, which is estimated from the AIA 94~\AA images (Section~1.2):
$$
U_{th} = 3k_BT\sqrt{EMVf}
$$
where is the Boltzmann constant, V is the volume of the flare region, is the filling factor of the flare region, which is commonly taken to be equal to one (Saint-Hilaire and Benz~2005; Emslie et al.~2012), suggesting that there is no fragmentation of the heating region inside the observed sources.

In our model, we consider the high-temperature sh region of the flare loops that we identify with the region of acceleration (or injection of nonthermal electrons into the loop) and the lower-temperature h region where the plasma is heated by nonthermal electrons and the heat flow from the sh region. To estimate the volume of each of these regions, we will use the pressure balance at the boundary of the sh and h regions:
$$
n_hk_BT_h + \frac{\rho_h v_h^2}{2} = n_{sh}k_BT_{sh} + \frac{\rho_{sh} v_{sh}^2}{2} + L_{sh}\rho_{sh}g_{sun}C
$$
Here, $g_{sun}$ is the surface gravity on the Sun, $L_{sh}$ is the length of the part of the loop in the sh region, and $Ñ<1$ is the coefficient that takes into account the correction for the projection of the gravity vector onto the magnetic field lines. The plasma number and mass densities in the h and sh regions are the sums of the electron and proton ones:$n_{h,sh}=n^e_{h,sh}+n^p_{h,sh}$ è $\rho_{h,sh}=n^e_{h,sh}m_e+n^p_{h,sh}m_p$, where $m_e$ and $m_p$ are the electron and proton masses. The formula described above is valid under the condition of quasi-stationarity and, accordingly, a small change in the plasma momentum in time $dt=4$ s. The disturbances in a magnetized plasma (in a flare loop) are transmitted with a magnetosonic velocity $v \approx\sqrt{c_s^2+v_A^2}\sim 10^8$ cm/s in a characteristic time $L/v\sim10$ s $>$ dt, $L\sim10^9$ cm. It follows from these estimates that the quasi-stationary regime may be considered on the time scales $dt$. Assuming subsonic plasma flow velocities and taking into account the characteristic plasma densities ($\lesssim 10^{11}$ cm$^{-3}$) and temperatures ($\sim 10^7$ K), we may neglect the terms in the momentum conservation law related to the gravity and the plasma flow pressure. Since the electron number density $n_e=\sqrt{EM/Vf}$, we have the relation
$$
T_h\sqrt{\frac{EM_h}{f_h}} = T_{sh}\sqrt{\frac{EM_{sh}}{f_{sh}}} \mbox{ ,}
$$
where $f_{sh}$ and $f_h$ are the volume fractions of the flare region (filling factors) occupied by the super-hot and hot plasmas. Assuming that the heating occurs in the entire flare volume (with the filling factor $f=1$), we have $f_h+f_{sh}=f=1$ and obtain the relation
$$
\alpha = \frac{f_{sh}}{f_{h}}=\frac{EM_{sh}T_{sh}^2}{EM_{h}T_{h}^2}
$$
whence $f_{sh}=f\alpha/(1 + \alpha)$ and $f_h = f/(1 + \alpha)$. The volumes of the sh and h regions are estimated as $V_{sh} = f_{sh}V$ and $V_h = f_hV$, respectively. The calculated time dependence of $f_{sh}$ is presented in Fig.\,7B. An increase in the filling factor of the very hot plasma is observed in the first minute, which is followed by its very slow decrease to the end of the investigated time interval.

\subsection{Calculating the Energy of the Nonthermal Electrons}

The energy contained in the nonthermal electrons is calculated from the following formula (Fletcher et al.~2007) within the thick-target model (Brown~1971):
$$
P_{nonth}(E>E_{low}) = 5.3\times 10^{24}\gamma^2(\gamma-1)\beta(\gamma - \frac{1}{2},\frac{3}{2})AE_{low}^{1-\gamma}\mbox{~   [erg s}^{-1}\mbox{]}
$$
This formula is used to calculate the energy of the nonthermal electrons from a power-law HXR spectrum of the form $I(E) = AE^{\gamma}$ with a low-energy cutoff of the nonthermal electron spectrum $E_{low}$; the power-law index for the nonthermal electrons is, accordingly, $\delta=\gamma-1$. At the very beginning of the impulsive phase, the SXR sources were observed at the chromospheric footpoints of the flare loops (Fig.\,ÇÀ). In this case, the thick-target approximation can traditionally be used. Subsequently, the SXR sources were observed already in the coronal part of the loops (Figs.\,3B and 3C). In this case, the application of a thick target is justified by the fact that the electrons with energies below $\sqrt{3C_{coll}n_h^pL_h}\sim 80$ keV (Brown et al.~1973; $C_{coll}=3.64\times10^{-18}$ keV$^2$ cm$^2$) lose their energy while passing through a plasma column corresponding to the loop half-length. In this case, the thick-target model can be used (Veronig and Brown~2004).

The most important parameter that determines the energy of the nonthermal electrons is the low-energy boundary of their spectrum $E_{low}$. It is considered in the simplest case and most commonly as a cutoff of the spectrum, i.e., it is assumed that there are no nonthermal particles for $E<E_{low}$. It is very difficult to determine $E_{low}$ from the observational data, because the thermal part of the X-ray spectrum masks the peculiarities of the nonthermal component. The low-energy boundary is commonly assumed to be equal to $\approx$15-30 keV, which is a formal choice, or the low-energy cutoff is considered as an additional variable parameter in the least-squares method when fitting the X-ray spectra. In reality, the shape of the low-energy part of the spectrum for the nonthermal electrons is determined by the physics of their acceleration. Here, we provide an estimate of the low-energy boundary based on the model presented in Section~4.1.

We assume that the electron acceleration occurs in the sh region of the flare loops (or immediately above it), with the minimum (critical) energy in the spectrum of accelerated electrons $E_c$ at the exit from the sh region being determined in such a way that the electrons with energies $E<E_c$ lose their energy through Coulomb collisions when propagating through the sh region. This energy is estimated as $E_c=\sqrt{3C_{coll}n_{sh}^pL_{sh}}$, where $n_{sh}^p$ is the proton number density in the sh region. For $n_{sh}^p\approx 3\times10^{11}$ cm$^{-3}$ and $L_{sh}\approx8\times10^7$ cm, $E_c\approx20$ keV. Accordingly, we took this value as a fixed low-energy boundary of the spectrum for the nonthermal electrons escaping from the sh region: $E_{low} = E_c$.

\subsection{Energy Balance}

In this section, we will consider the energy balance between all the included channels of energy release both in the entire flare region and in the sh and h regions.

In addition to calculating the time derivative of the plasma internal energy and the kinetic power of the accelerated electrons, it is also necessary to take into account the radiative heat losses from the entire flare region. For an X-ray-emitting plasma, the heat losses are estimated as $L_{rad}=EM\times10^{-17.73}T^{-2/3}$ for flare temperatures (Rosner et al.~1978). The radiative losses in ultraviolet emission are calculated by integrating the spectrum taken with the EVE instrument in the wavelength range 6.5-40 nm. We exclude the radiative losses in other ranges of the electromagnetic spectrum from consideration. However, this should not have a strong effect on the final result, because, for example, Emslie et al.~(2012) showed that, within the error limits, the bolometric losses through radiation exceed insignificantly the losses through SXR emission.

Figure~7A presents the following time profiles: the time derivative of the internal energy from RHESSI data, the kinetic power of the nonthermal electrons, the radiative heat losses from GOES and EVE data, and our estimate of the heat flux from the sh region into the h region of the flare loops through thermal conduction ($L_{cond}$). This heat flux is calculated to estimate the heating of the h region through thermal conduction from the sh region. The thermal conduction is assumed to be classical: $L_{cond}\approx 4\times 10^{-6}T^{5/2}(Tsh - Th)/L$, where $T\approx(Tsh + Th)/2$, $L$ characterizes the length scale of the temperature gradient that is taken to be equal to the flare loop length.

The time derivative of the plasma internal energy in the sh region is, on average, a factor of 5 smaller than that for the h region. Beginning approximately from the second minute, plasma cooling is observed in the entire flare region, with this roughly corresponding to the beginning of dominance of the radiative losses over the plasma heating by nonthermal electrons. The total rate of radiative losses according to GOES and EVE is equal to the rate of decrease of the plasma internal energy in the flare region. Thus, the cooling phase ($dU_{th}/dt<0$) in the event under consideration is satisfactorily explained in terms of the chosen model of the flare region.

During the heating phase (the first two minutes), the time derivative of the plasma internal energy in the h region is, on average, larger than the kinetic power of the nonthermal electrons. It can be assumed that there is an additional heating source of the h region or that not only the electrons with energies higher than $E_c$ but also those with lower energies are involved in the acceleration process. As an additional heating source of the h region, we may consider the heat flux into it through thermal conduction from the sh region; its estimate is also presented in Fig.\,7A (gray histogram). The heat flux is seen to be at least an order of magnitude larger than the time derivative of the internal energy. This provides evidence for the fact that the thermal conduction was clearly overestimated and its classical description is unsuitable for this case. It may be necessary to consider suppressed anomalous thermal conduction (see the Discussion).

Figure~8 presents our calculations of the energy release in the case of applying the one-temperature approximation to the RHESSI spectra. In this case, to calculate the energies, we used the formulas described in preceding sections, with the only difference that the low-energy boundary of the nonthermal electron spectrum was a free parameter and its values were determined from the results of our analysis of the RHESSI spectra (Fig.\,4E2). It can be seen from Fig.\,8 that the kinetic power of the nonthermal electrons is much greater than the remaining channels of energy release, by more than an order of magnitude in some periods. The total energy of the nonthermal electrons, $\approx 4\times10^{32}$ erg, is higher than the total energy of the emission from GOES data, $\approx 5\times10^{30}$ erg, by two orders of magnitude. Since the energy release in a flare is connected with the magnetic energy, its value (lower boundary) can be estimated from the energy balance $B^2V/8\pi\sim\int_{t_{imp}} P_{nonth}dt$, where $t_{imp}$ is the duration of the impulsive phase. Hence we obtain $Â\sim10^3$ G. This is an overestimated value for the coronal magnetic field. Qiu et al.~(2009) and Fleishman et al.~(2013) estimated the magnetic field in the coronal part of flare loops from microwave observations to be $<150$ G. It can be seen from this estimate that the kinetic power of the nonthermal electrons was overestimated, which may suggest that the assumption about a nonthermal nature of the X-ray emission in the range 25-50 keV is artificial. The energy balance can be restored by assuming a higher value of the low-energy cutoff for the spectrum of accelerated electrons, but the RHESSI spectra are fitted with a large $\chi^2$ in this case.

In the case of a two-temperature fit to the RHESSI spectra, we obtain (by time integration) a total energy of the nonthermal electrons $\approx 2.5\times10^{30}$ erg, in agreement with the radiative heat losses according to GOES data $\approx 5\times10^{30}$ erg. The magnetic field is estimated from the energy balance $B^2V/8\pi\sim\int_{t_{imp}}P_{nonth}dt$ to be ~100 G, corresponding to the calculations in Qiu et al.~(2009) where a two-ribbon flare with a similar spatial structure, with an arcade of flare loops, was also investigated.

Below, we give our estimate of the energy release per unit time $Q$ with which the entire flare process (for example, magnetic reconnection) is associated. Neglecting the convective heat flow through chromospheric evaporation from the h region into the sh region, the change in mechanical energy, and the work of the pressure forces, we can write the energy balance equations for the h and sh regions:
$$
\dot{U_{th}^{sh}} = Q - L_{rad}^{sh}-L_{cond}-P_{nonth}
$$
$$
\dot{U_{th}^{h}} = P_{nonth} + L_{cond} - L_{rad}^{h}
$$
In the first equation, the minus sign in front of the kinetic power of the nonthermal electrons means that the acceleration occurs in the sh region; thus, the nonthermal electrons carry the energy away. Adding these two relations, we obtain the energy balance equation for the entire flare region in which the thermal conduction is disregarded, because it leads only to a redistribution of heat inside the flare region:
$$
\dot{U_{th}} = Q - L_{rad}
$$
Hence we can estimate the rate of energy release at maximum, $Q\sim10^{29}$ erg s$^{-1}$, and the total energy release in the time of the impulsive phase, $\sim 10^{31}$ erg. Figure~6 (panel~D) presents the ratio $Q/P_{nonth}$ that shows that the kinetic power of the accelerated electrons is, on average, smaller than the total energy release by a factor of 5.

\section{DISCUSSION}

Caspi and Lin~(2010) investigated the X-ray spectra of the July 23,~2002 flare in the two-temperature approximation. As a result, they obtained approximately the same temperatures for the hot ($\sim18$ MK) and super-hot ($\sim 45$ MK) plasmas as those for the August 9,~2011 event. Caspi and Lin assumed that the super-hot plasma was concentrated in the loops produced by magnetic reconnection in the corona, while the h-region was filled with the plasma ejected from the dense atmosphere (chromospheric evaporation) as a result of its heating by nonthermal electrons. The plasma in the flare region is heated in two different ways: (1) by accelerated electrons interact¬ing with the plasma and (2) by thermal conduction through the energy going from the region of primary energy release (electron acceleration). In this case, high coronal magnetic fields ($\sim 100$ G) are needed for the magnetic loops with a plasma at a very high temperature to be in equilibrium.

To interpret our observations and the results of our spectral analysis, we used the model described in Section~3.1, where the high-temperature region (sh region) is assumed to be located in the coronal part of the flare loops above the h region (or the magnetic cusp) and to be connected with the electron acceleration and the main process of energy release. The X-ray images of the flare region for the July 23,~2002 and August 9,~2011 events confirm the spatial structure in the model described in Section~3.1. However, a clear shift between the centers of the X-ray sources corresponding to the emission from the h- and sh-regions was observed in the first event, because the linear scales of the loops in this event were larger than those in the August 9, 2011 event.

We consider the two-temperature approximation, but, in reality, the temperature distribution in the flare region (the differential emission measure, DEM) is continuous and must be analyzed by the methods described, for example, in Fludra and Sylwester~(1985) and Hannah and Kontar~(2012). It is worth noting that the present-day DEM calculations do not touch on the range of very high flare temperatures ($>20$ MK) or the calculations have large errors. It follows from our results and the conclusions reached by Caspi and Lin~(2010) that the plasma with temperatures above the arbitrary boundary of 30 MK is connected with the processes occurring in the region of primary energy release. Thus, the DEM calculations available to date allow only the gasdynamic processes to be investigated and, having begun our investigation of the temperature distribution in the flare region above 30 MK, we will be able to trace the evolution of the plasma associated with magnetic reconnection and acceleration processes (Caspi et al. 2014).

The applied two-temperature model of the flare region with acceleration in the sh-region (or immediately above it) suggests the existence of a heat flow into the h-region, along with the heating through nonthermal electrons. The presence of such an additional heating source in the h-region is also justified by the results of our calculations of the rate of total energy release $Q$, which turned out to be higher than the kinetic power of the nonthermal electrons (Fig.\,7B). Our estimates show that when using the classical thermal conductivity, we have an excessively large heat flow going from the sh-region into the h-region that disagrees with the remaining channels of energy release. The thermal conduction is most likely to be anomalous (Oreshina and Somov 2011a, 2011b) and related to magnetoplasma turbulence. In addition, the nonthermal electrons themselves can be responsible for the excitation of ion-acoustic turbulence that leads to the suppression of anomalous thermal conduction. This scenario was discussed, for example, in Astrelin et al.~(1998) as applied to the suppression of longitudinal plasma thermal conduction by a beam of energetic electrons injected into a laboratory magnetic trap.

The anomalously high flare plasma temperature measured with GOES can also be circumstantial evidence for anomalous thermal conduction in the flare region considered. As we showed in Section~3, the high temperature can be connected with a small amount of plasma at a ``low'' temperature ($<20$ MK). The absence (shortage compared to normal flares) of a low-temperature component can be related to anomalous thermal conduction, which prevents the rapid plasma cooling in the flare region and the outflow of heat into other regions of the flare volume.

The zone of a super-hot ($\sim 45$ MK) plasma may have been observed in the flare loops under consideration precisely because of the presence of anomalous thermal conduction. Strong heating, which increases the plasma $\beta$ to $\sim 1$, occurs through the process of initial heating. Turbulence can develop in such a plasma, causing the temperature to rise through a reduction in the removal of heat due to anomalous thermal conduction. In other words, the conditions for overheating of the flare region are realized for characteristic times $\tau_{cond}\gg\tau_{heat}$. However, this process is possible only under suppressed (anomalous) thermal conduction. The sh region can be heated by the low-energy electrons with energies below $E_c=20$ keV forming in a less dense plasma, which are then completely thermalized in the sh region through Coulomb collisions. For example, consider nonthermal electrons with a power-law spectrum with boundaries of 10 and 20 keV. For the parameters of the spectrum presented in Fig.\,4, we obtain $P_{nonth}$(10 keV $<E<$ 20 keV)$\approx1.7\times 10^{28}$ erg s$^{-1}$, which is comparable to $P_{nonth}$($E>$20 keV)$\approx 2.0\times 10^{28}$ erg s$^{-1}$. Thus, a substantial energy is contained in the low-energy electrons that can heat the sh region. However, other heating mechanisms, for example, the dissipation of electric currents or shock waves, are also possible. It is worth noting that here we disregarded the possible plasma heating by accelerated protons, although Emslie et al.~(2012) showed that the energy of the accelerated ions could be comparable to the energy of the nonthermal electrons in some events.

The existing models of gasdynamic processes in a flare region (see, e.g., Fisher et al.~1985; Boiko and Livshits~1995) used the classical thermal conduction. In light of the presented reasoning, allowance for the anomalous thermal conduction in gasdynamic calculations can be of special interest for future works. However, first, additional experimental confirmations of the possibility of producing the anomalous thermal conduction in flare regions are needed. This requires performing an analysis similar to that presented here at least for several other flares. Second, it is necessary to clarify the nature of the anomalous thermal conduction for the problem to be self-consistent. For example, if the particles generate waves, then the plasma—kinetic processes for beams of nonthermal electrons should be taken into account in the calculations.

Under conditions of a medium with a plasma $\beta\sim 1$, it is most likely necessary to take into account the convective heat transfer inside the sh region and, hence, the filling factor in such events must be close to unity, which gives a justification for our calculations where $f=1$. Caspi et al.~(2014) also provide arguments that the filling factor is close to unity.

\section{CONCLUSIONS}

(1)	According to GOES, the August 9,~2011 solar flare is characterized by an anomalously high temperature (32.5 MK at the peak). This is the ``hottest'' GOES flare over the period from 2000 to 2012. The
most likely cause of the high temperature calculated from GOES data is a small amount of plasma with relatively low temperatures ($T<20$ MK) compared to ``normal'' flares in which the temperature derived from GOES data is $<30$ MK.

(2)	The RHESSI X-ray spectra were fitted in the two-temperature and one-temperature approximations at energies below 50 keV and by a power law at energies above 50 keV. The two-temperature model describes better the energy release from the viewpoint of energy balance in the flare region.

(3)	The plasma heating to super-hot temperatures is most likely associated with the primary energy release and the suppression of thermal conduction in the flare loops. The super-hot region is characterized by a high plasma density ($1-3\times 10^{11}$ cm$^{-3}$), a plasma $\beta\sim 1$, and a fraction of the total flare volume equal to $\approx 0.10-0.16$.

(4)	The anomalous (suppressed) thermal conduction can be a very important factor affecting the plasma internal energy distribution in a flare region. The anomalous thermal conduction can be both a consequence of the generation of plasma waves during the propagation of charged particles and a consequence of developed turbulence in a super-hot magnetized plasma. In future models of gasdynamic processes in a flare region, it will possibly be necessary to take into account the emergence of anomalous thermal conduction. However, additional observations are needed to confirm (or rule out) the
suppression of thermal conduction in flare regions. Additional theoretical studies of the thermal conduction in the nonequilibrium plasma of flare regions are also needed.

(5)	A further study of the differential emission measure for the flare plasma at temperatures above 30 MK can help in understanding the energy release and the acceleration of charged particles during flares and the lower-temperature regions in understanding the gasdynamic processes in flare loops.

\section*{ACKNOWLEDGMENTS}

This work was financially supported in part by the Russian Foundation for Basic Research (project no. 13-02-91165), Presidential grant MK-3931.2013.2, Program P-22 of the Presidium of the Russian Academy of Sciences, and ``Support of Young Scientists''. We are grateful to the referee for the careful reading and the remarks.

\section*{REFERENCES}

\begin{enumerate}
\item Astrelin V.T, Burdakov A.V., Postupaev V.V. Plasma Physics, 24, 450-462 (1998)
\item Boiko A.Ya., Livshits M.A. Astron. Journal, 72, 381 (1995)
\item Brown J.C. Solar Physics, 18, 489-502 (1971)
\item Brown J.C. Solar Physics, 31, 143 (1973)
\item Veronig A.M., Brown, J.C. Astrophys. J., 603, 117-120 (2004)
\item Woods T.N., Eparvier F.G., Hock R. Solar Physics, 275, 115-143 (2012)
\item Den O.G., Somov B.V. Soviet Astronomy, 33, 149 (1989)
\item Dere K.P., Landi E., Young P.R., Del Zanna G., Landini M., Mason H.E. Astronomy \& Astrophysics, 498, 915-929 (2009)
\item Emslie A.G., Dennis B.R., Shih A.Y., Chamberlin P.C., Mewaldt R.A., Moore C.S., Share G.H., Vourlidas A., Welsch B.T. Astrophys. J., 759, 18 (2012)
\item Caspi A., Lin R.P. Astrophys. J. Letters,725, 161-166 (2010)
\item Caspi A., Krucker S., Lin R.P. Astrophys. J., 781, 11 (2014)
\item Caspi A., McTiernan, J.M., Warren H.P. Astrophys. J., 788, 6 (2014)
\item Kovalev V.A., Chernov G.P., Hanaoka I. Astronomy Letters, 27, 267-275 (2001)
\item Qiu J., Gary D.E., Fleishman G.D. Solar Physics, 255, 107-118 (2009)
\item Lemen J.R., Title A.M., AkinDavid J. et al. Solar Physics, 275, 17-40 (2012)
\item Lin R.P., Dennis B.R., Hurford G.J., Smith D.M. et al. Solar Physics, 210, 3-32 (2002)
\item Liu S., Li Y., Fletcher L. Astrophys. J., 769, 10 (2013)
\item MacNeice P., Burgess A., McWhirter R.W.P., Spicer D.S. Solar Physics, 90, 357-382 (1984)
\item Oreshina A.V. and Somov B.V. Astronomy Letters, 37, 726-736 (2011)
\item Oreshina A.V. and Somov B.V. Astronomy Letters, 66, 286-291 (2011)
\item Ryan D.F., Milligan R.O., Gallagher P.T., Dennis B.R., Tolbert A.K, Schwartz R.A., Young C.A. Astrophys. J., 202, 15 (2012)
\item Rosner R., Tucker W.H., Vaiana G.S. Astrophys. J., 220, 643-665 (1978)
\item Svestka Z., Poletto G. Solar Physics, 97, 113-129 (1985)
\item Saint-Hilaire P. and Benz A.O. Astronomy \& Astrophysics, 435, 743-752 (2005)
\item Somov B.V., Kosugi T. Astrophys. J., 485, 859 (1997)
\item Thomas R.J., Starr R. and Crannell C.J. Solar Physics, 95, 323-329 (1985)
\item Fisher G.H., Canfield R.C., McClymont A.N. Astrophys. J., 289, 425 (1985)
\item Fleishman G.D., Kontar E.P., Nita G.M., Gary D.E. Astrophys. J., 768, 13 (2013)
\item Fletcher L., Hannah I.G., Hudson H.S., Metcalf T.R.  Astrophys. J., 656,1187-1196 (2007)
\item Fludra A., Sylwester J. Solar Physics, 105, 323-337 (1986)
\item Hannah I.G., Kontar E.P) Astronomy \& Astrophysics, 539, 14 (2012)
\item Holman G.D. Astrophys. J., 586, 606-616 (2003)
\end{enumerate}

\clearpage

\begin{figure}
\epsscale{.80}
\plotone{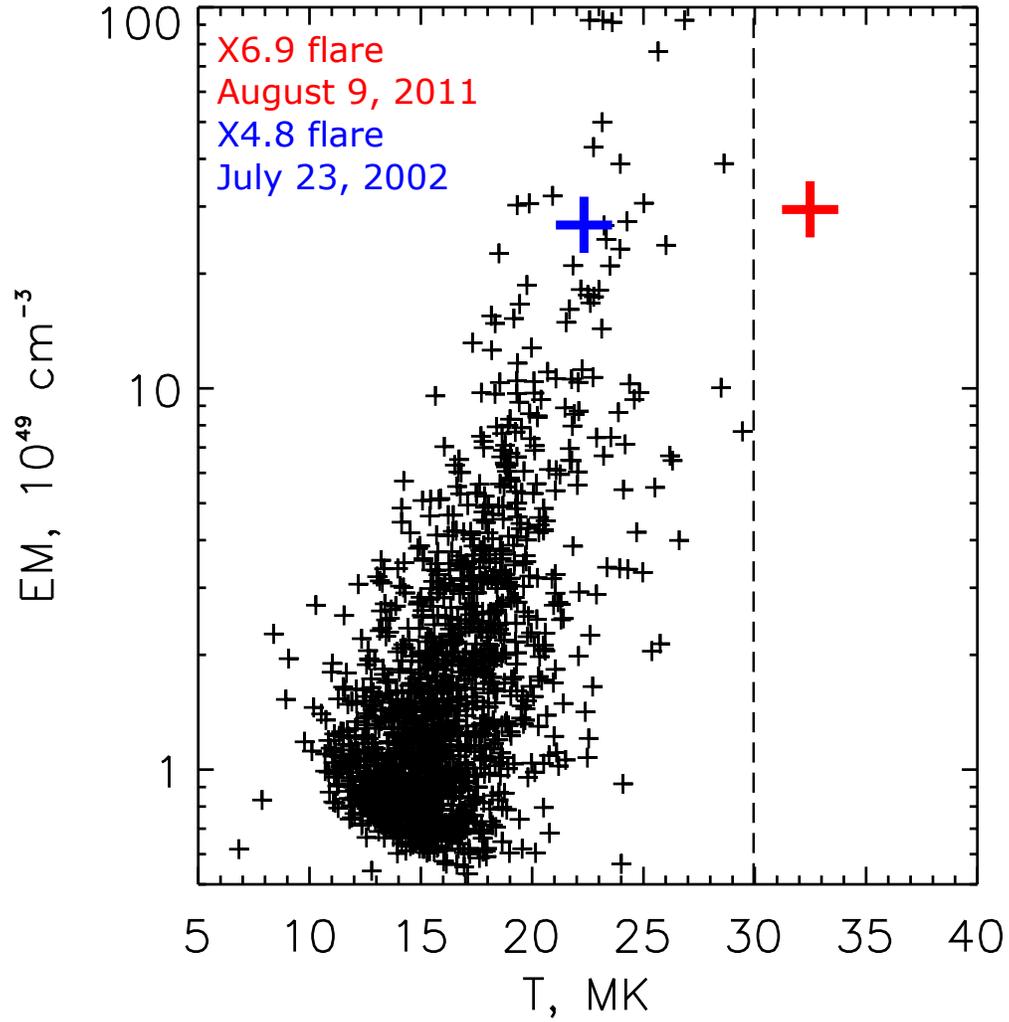}
\caption{Peak emission measure-peak temperature diagram for M- and X-class flares over the period 2000-2012 from GOES data. The red and blue crosses mark the August 9, 2011 and July 23, 2002 flares, respectively. \label{fig1}}
\end{figure}

\clearpage
\begin{figure}
\epsscale{1.0}
\plotone{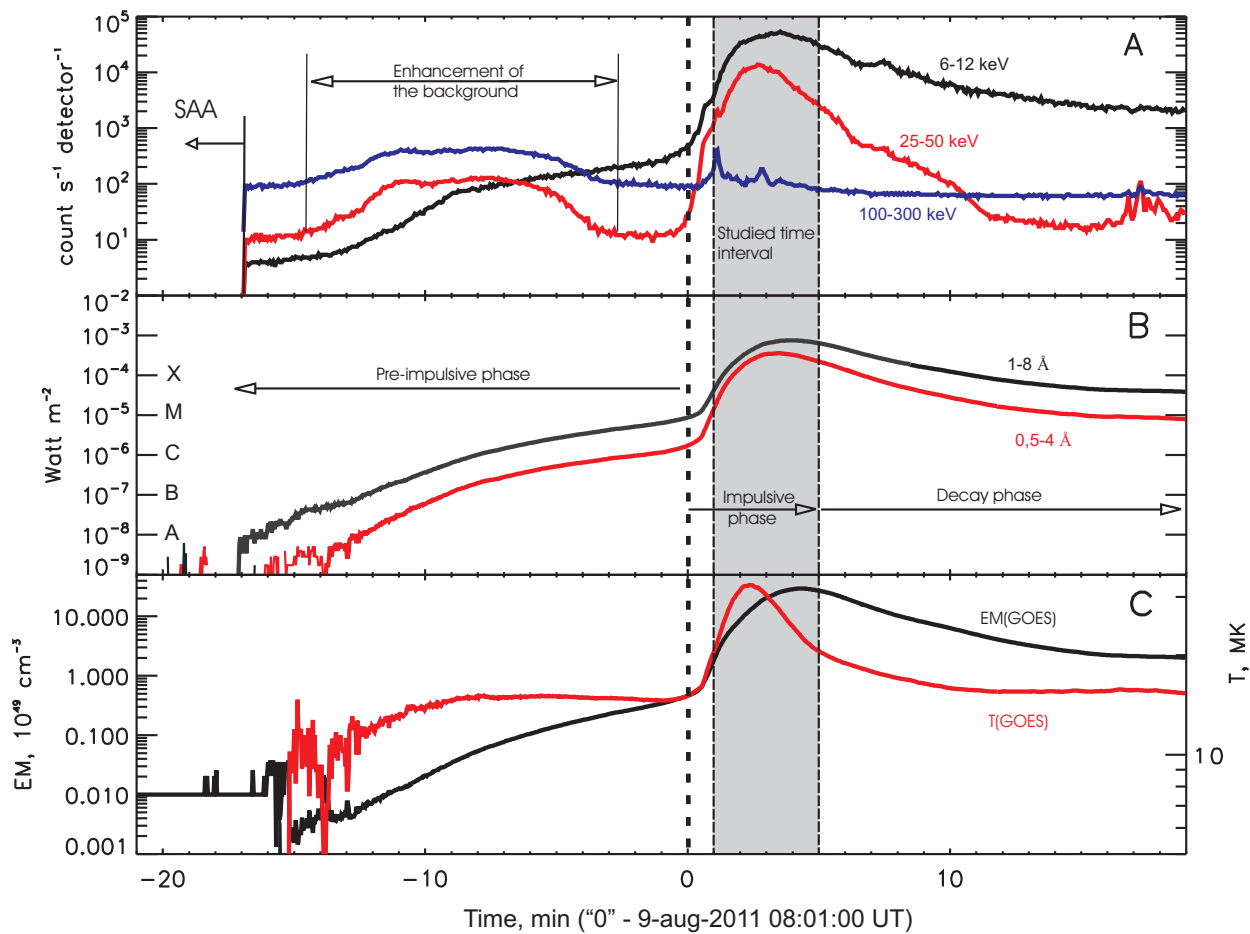}
\caption{Time profiles: (A) for the RHESSI count rate in three energy ranges, 6-12, 25-50, and 100-300 keV; (B) for the X-ray flux from GOES data in two channels, 0.5—4 and 1—8 A; (C) for the emission measure and temperature calculated from GOES observations. The gray band designates the time interval in which the energy release was analyzed in detail; the arrows mark the characteristic periods of the flare. \label{fig2}}
\end{figure}

\clearpage
\begin{figure}
\epsscale{1.00}
\plotone{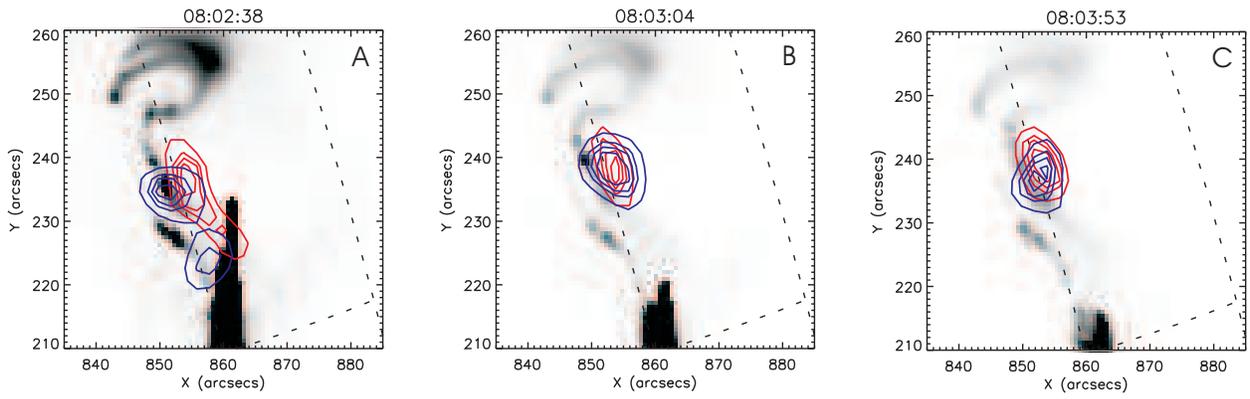}
\caption{Substrate: the AIA images in the 94 \AA channel. The contours of the X-ray images correspond to 50, 70, 80, and 90 \% of the intensity in the brightest pixel. The red and blue colors are for 3—15 and 60-200 keV, respectively (RHESSI data). \label{fig3}}
\end{figure}

\clearpage
\begin{figure}
\epsscale{1.00}
\plotone{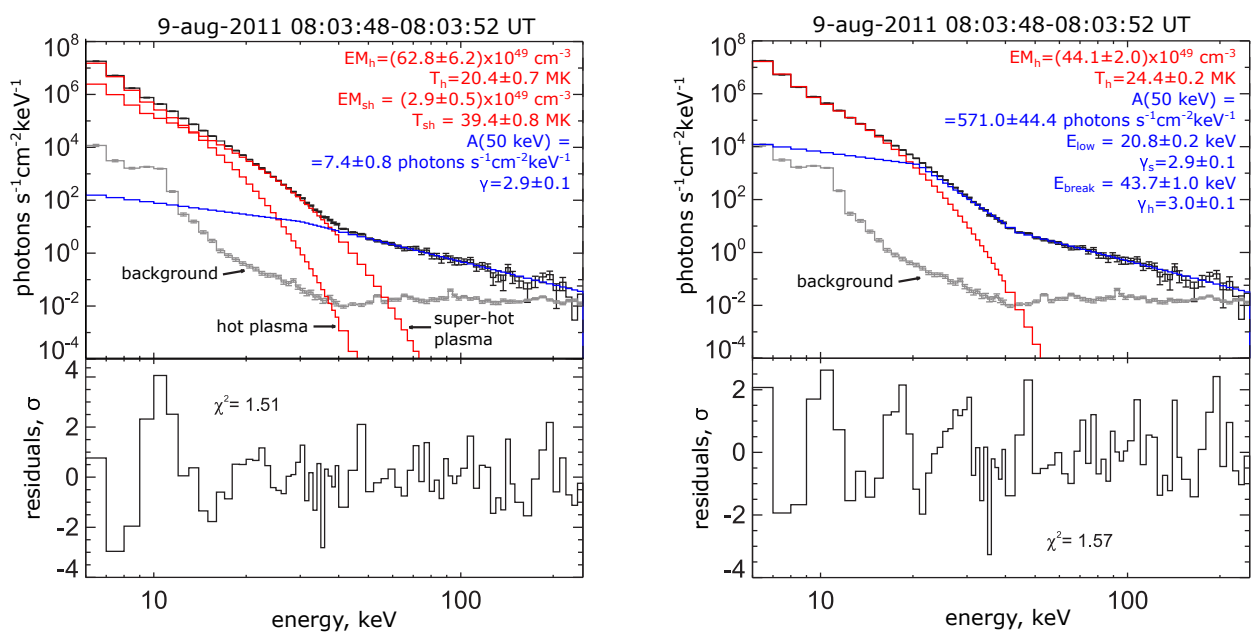}
\caption{Left panel — an example of the two-temperature fit to the RHESSI X-ray spectrum. Right panel — an example of the one-temperature fit to the RHESSI X-ray spectrum. The $\chi^2$ fit is presented in the lower part of the figure. \label{fig4}}
\end{figure}

\clearpage
\begin{figure}
\epsscale{1.00}
\plotone{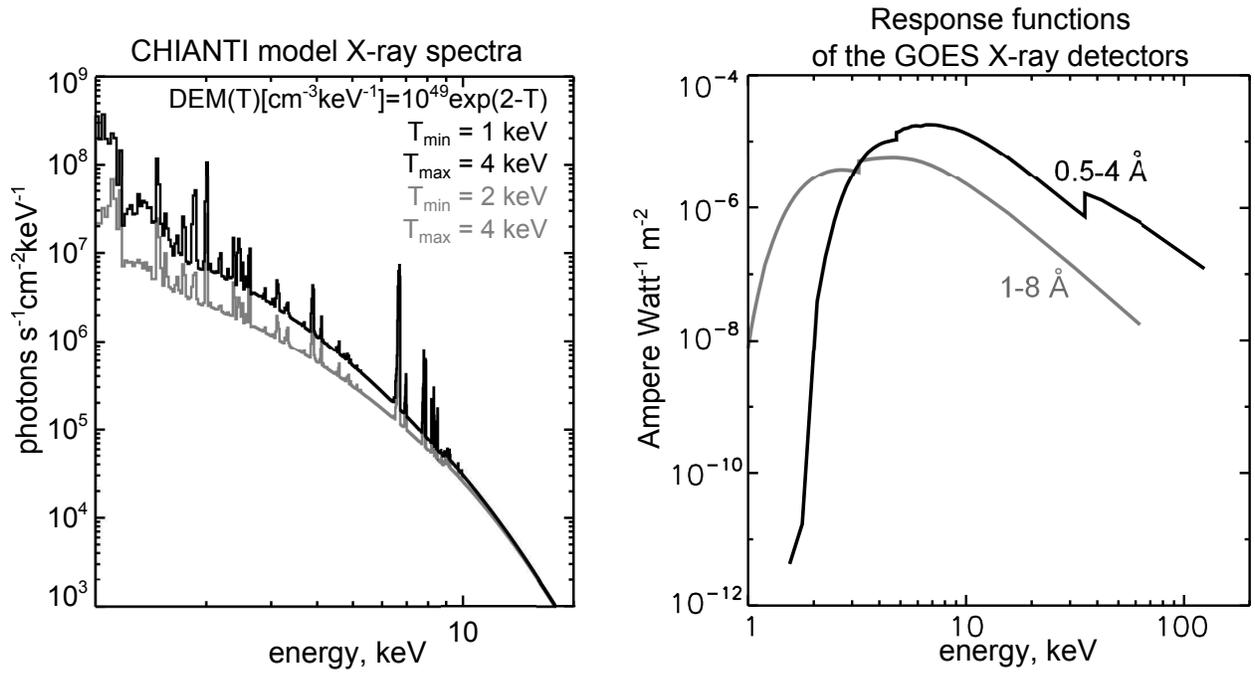}
\caption{Left panel — the model X-ray spectra of the coronal plasma obtained by using the CHIANTI database for two differential emission measures (indicated in the figure). Right panel — the response functions of the GOES X-ray detectors to the emission being recorded. \label{fig5}}
\end{figure}

\clearpage
\begin{figure}
\epsscale{1.00}
\plotone{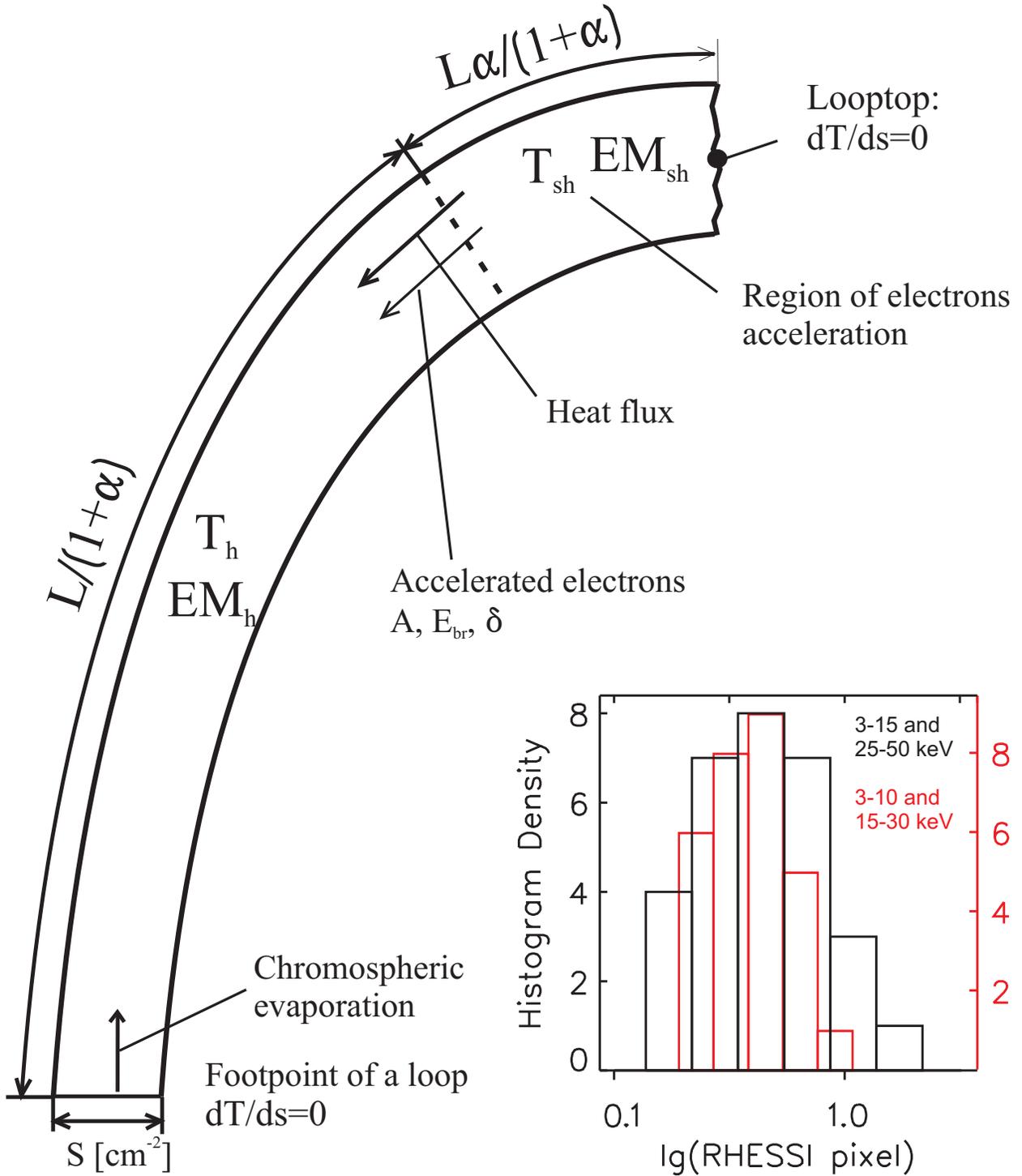}
\caption{Schematic model of the flare region: the histogram presents the distribution of centroid separations for the SXR sources in two different energy ranges: 3—15 and 25—50 keV (black), 3—10 and 15—30 keV (red). \label{fig6}}
\end{figure}

\clearpage
\begin{figure}
\epsscale{1.00}
\plotone{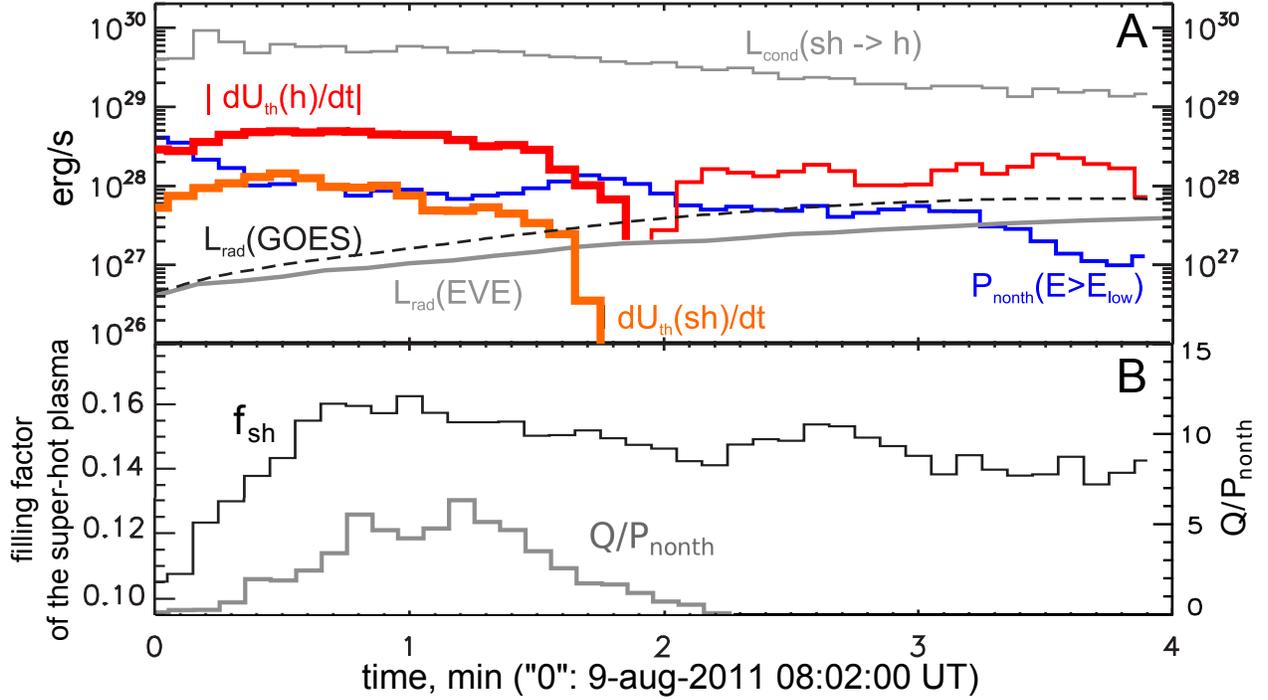}
\caption{ Results of our calculations of the energetics within the two-temperature model (Section~3.1). Panel A: red thick and red thin histograms show the positive and negative (absolute value) time derivatives of the plasma internal energy in the h region; red thick histogram shows the time derivative of the plasma internal energy in the sh region; blue thin histogram shows the kinetic power of the nonthermal electrons $E>E_{low}$; the black dashed line indicates the radiative losses from GOES data; the gray line indicates the radiative losses from EVE data; the gray thin histogram shows the heat flux from the sh region into the h region. Panel B: the black line indicates the filling factor of the sh region (Section~3.2) and the gray line indicates the ratio of the total energy release per unit time ($Q$) to the kinetic power of the accelerated electrons ($P_{nonth}$). \label{fig7}}
\end{figure}

\clearpage
\begin{figure}
\epsscale{1.00}
\plotone{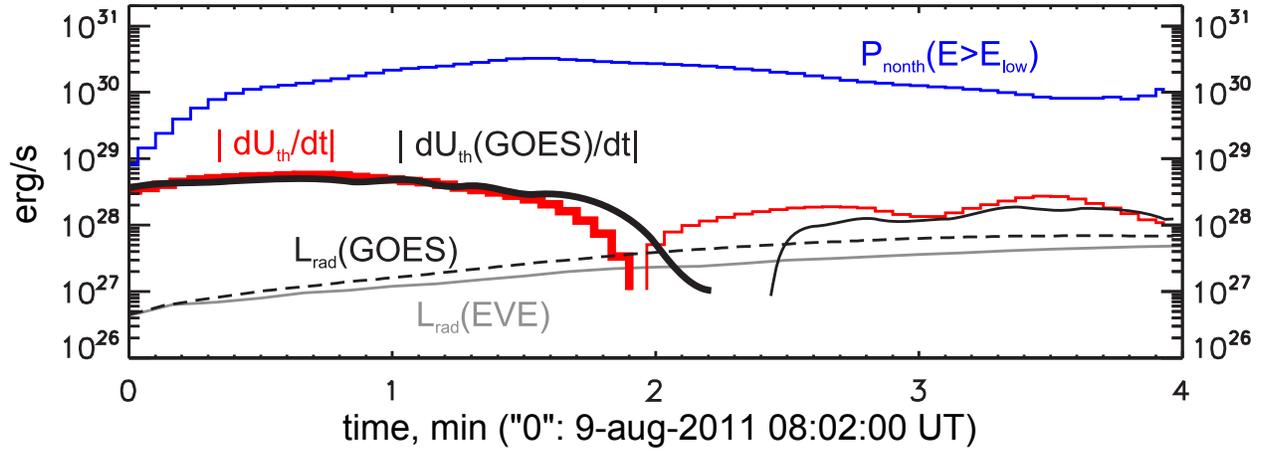}
\caption{Results of our calculations of the energetics in the one-temperature approximation. Red thick and thin histograms show the positive time derivative and the absolute value of the negative time derivative of the plasma internal energy from RHESSI data; blue histogram shows the kinetic power of the accelerated electrons from RHESSI data; the black dashed line indicates the radiative heat losses in SXR emission from GOES data; the radiative heat losses in UV emission from EVE data. \label{fig8}}
\end{figure}

\clearpage
\end{document}